\documentstyle[12pt,epsfig]{article}
\begin{document}

\begin{flushright}
\framebox{\bf hep-ph/0107257} \\
\end{flushright}

\vspace{0.5cm}

\begin{center}
{\Large\bf Determining the Factorization Parameter and \\
Strong Phase Differences in $B\rightarrow D^{(*)}\pi$ Decays}
\end{center}

\vspace{0.1cm}

\begin{center}
{\bf Zhi-zhong Xing}
\footnote{Electronic address: xingzz@mail.ihep.ac.cn} \\
{\it Institute of High Energy Physics, P.O. Box 918 (4),
Beijing 100039, China}
\end{center}

\vspace{2.5cm}

\begin{abstract}
The first observation of the color-suppressed decay mode
$\bar{B}^0_d \rightarrow D^{(*)0} \pi^0$ by the Belle and CLEO
Collaborations makes a quantitative analysis of the isospin relations
for the amplitudes of $B\rightarrow D^{(*)}\pi$ possible.
The strong (isospin) phase difference in $B\rightarrow D\pi$ transitions
is found to be about $29^\circ$ by use the Belle data or $26^\circ$ by
use of the CLEO data, implying that final-state interactions
might not be negligible. Applying the factorization approximation to
$I=3/2$ and $I=1/2$ isospin amplitudes of $B\rightarrow D\pi$ decays,
we obtain the ratio of the effective Wilson coefficients
$a^{\rm eff}_1$ and $a^{\rm eff}_2$:
$a^{\rm eff}_2/a^{\rm eff}_1 \approx 0.27$. A similar analysis shows
that the magnitude of final-state interactions in $B\rightarrow D^*\pi$
might be comparable with that in $B\rightarrow D\pi$,
and the factorization hypothesis works consistently in both of them.
\end{abstract}

\newpage

Two-body nonleptonic decays of the type $B\rightarrow D\pi$
have been of great interest in $B$ physics for
a stringent test of the factorization hypothesis and a
quantitative analysis of final-state interactions \cite{Stech}.
The color-suppressed decay mode $\bar{B}^0_d \rightarrow D^0 \pi^0$
has for the first time been observed by the Belle \cite{Belle} and
CLEO \cite{CLEO} Collaborations. Its branching ratio is found to be
\begin{equation}
{\cal B}_{00} \; = \; \left \{ \matrix{
(2.9^{+0.4}_{-0.3} \pm 0.6 ) \cdot 10^{-4} ~~~~ ({\rm Belle}) \; ,
\;\;\;\;\;\; \cr
(2.6 \pm 0.3 \pm 0.6 ) \cdot 10^{-4} ~~~~ ({\rm CLEO}) \; . } \right .
%		(1)
\end{equation}
In comparison, the branching ratios of the color-favored decay modes
$\bar{B}^0_d \rightarrow D^+ \pi^-$ and
$B^-_u \rightarrow D^0\pi^-$ are \cite{PDG}
\begin{eqnarray}
{\cal B}_{+-} & = & \left (3.0 \pm 0.4 \right ) \cdot 10^{-3} \; ,
\nonumber \\
{\cal B}_{0-} & = & \left (5.3 \pm 0.5 \right ) \cdot 10^{-3} \; .
%		(2)
\end{eqnarray}
One can see that the naively-expected color suppression {\it does}
appear for $\bar{B}^0_d \rightarrow D^0 \pi^0$; i.e.,
${\cal B}_{+-} \sim {\cal B}_{0-} \sim 3^2 \cdot {\cal B}_{00}$.
With the help of the new experimental data in Eq. (1), one is now
able to analyze the isospin relations for the amplitudes of
$B\rightarrow D\pi$ decays in a more complete way than before
(see, e.g., Refs. \cite{Yamamoto,Xing95,Suzuki}). Then it becomes possible
to check whether final-state interactions are significant in such
exclusive $|\Delta B| = |\Delta C| =1$ transitions, and
whether the factorization approximation works well.

\vspace{0.3cm}

The main purpose of this paper is to determine the strong (isospin)
phase difference and the factorization parameter
$a^{\rm eff}_2/a^{\rm eff}_1$ in $B\rightarrow D\pi$ decays.
The former is found to be about $29^\circ$ by use of the Belle data
or $26^\circ$ by use of the CLEO data, implying that final-state
interactions might not be negligible. We obtain
$a^{\rm eff}_2/a^{\rm eff}_1 \approx 0.27$, a value in good agreement
with the theoretical expectation. At the end of this paper, we present
a similar isospin analysis for the decay modes
$B\rightarrow D^*\pi$. Our result shows that the magnitude of
final-state interactions in $B\rightarrow D^*\pi$ transitions might
be comparable with that in $B\rightarrow D\pi$ transitions, and the
factorization approximation works consistently in both of them.

\vspace{0.3cm}

The effective weak Hamiltonian responsible for
$\bar{B}^{0}_{d}\rightarrow D^{+}\pi^{-}$,
$\bar{B}^{0}_{d} \rightarrow D^{0}\pi^{0}$ and
$B^{-}_{u}\rightarrow D^{0}\pi^{-}$ transitions \cite{Buras}
has the isospin configuration $|1, -1\rangle$.
Therefore their amplitudes, defined
respectively as $A_{+-}$, $A_{00}$ and $A_{0-}$, can be
decomposed as follows \cite{Xing95}:
\begin{eqnarray}
A_{+-} & = & A_{3/2} ~ + ~ \sqrt{2} A_{1/2} \; ,
\nonumber \\
A_{00} ~ & = & \sqrt{2} A_{3/2} ~ - ~ A_{1/2} \; ,
\nonumber \\
A_{0-} & = & 3A_{3/2} \; ,
%		(3)
\end{eqnarray}
where $A_{3/2}$ and $A_{1/2}$ stand respectively for
$I=3/2$ and $I=1/2$ isospin amplitudes. In obtaining Eq. (3),
we have assumed that there is no mixture of $B\rightarrow D\pi$ with other
channels \cite{BSW}. It is obvious that three transition amplitudes
form an isospin triangle in the complex plane:
$A_{+-} + \sqrt{2} A_{00} = A_{0-}$.
Of course, the sizes of $A_{+-}$, $A_{00}$ and $A_{0-}$ can
straightforwardly be determined from the branching ratios given in
Eqs. (1) and (2). Then we are able to extract the ratio $A_{3/2}/A_{1/2}$,
both its size and its phase, by use of Eq. (3). We find
\begin{eqnarray}
r & \equiv & \left | \frac{A_{3/2}}{A_{1/2}} \right |
\nonumber \\
& = & \sqrt{\frac{{\cal B}_{0-}}
{3 \kappa ({\cal B}_{+-} + {\cal B}_{00}) - {\cal B}_{0-}}} \;\; ,
\nonumber \\ \nonumber \\
\delta & \equiv & \arg \left (\frac{A_{3/2}}{A_{1/2}} \right )
\nonumber \\
& = & \arccos \left [ \frac{ 3 \kappa ({\cal B}_{+-} - 2 {\cal B}_{00})
+ {\cal B}_{0-}}
{\sqrt{8 {\cal B}_{0-} [3 \kappa ({\cal B}_{+-} + {\cal B}_{00}) -
{\cal B}_{0-}]}} \right ] \; ,
%		(4)
\end{eqnarray}
where $\kappa \equiv \tau_{B^-_u}/\tau^{~}_{\bar{B}^0_d}
= 1.073 \pm 0.027$ \cite{PDG} measures
the difference between the life time of $\bar{B}^0_d$ and that of
$B^-_u$. On the other hand, the tiny phase-space corrections induced
by the mass differences $m^{~}_{D^{0}}-m^{~}_{D^{-}}$ and
$m_{\pi^{0}}-m_{\pi^{-}}$ have been neglected in obtained Eq. (4).

\vspace{0.3cm}

Using the central values of ${\cal B}_{+-}$, ${\cal B}_{00}$,
${\cal B}_{0-}$ and $\kappa$, we obtain the following result
for $r$ and $\delta$:
\begin{eqnarray}
r & \approx & \left \{ \matrix{
1.0 ~~~~ ({\rm Belle}) \; , \;\; \cr
1.0 ~~~~ ({\rm CLEO}) \; ;} \right .
\nonumber \\
\delta & \approx & \left \{ \matrix{
29^\circ ~~~~ ({\rm Belle}) \; , \;\; \cr
26^\circ ~~~~ ({\rm CLEO}) \; .} \right .
%		(5)
\end{eqnarray}
If errors of the input parameters are taken into account, we find that
$r$ may change from 0.5 to 1.5 and $\delta$ can be as small
as $0^\circ$ in the extreme case.
It is most likely, however, that $\delta$ takes the value given in
Eq. (5), implying that final-state interactions in $B\rightarrow D\pi$
transitions might not be small. In addition, $r\approx 1$ means that
the two isospin amplitudes have comparable contributions to the
decay modes $\bar{B}^0_d\rightarrow D^+\pi^-$ and
$\bar{B}^0_d\rightarrow D^0\pi^0$.

\vspace{0.3cm}

Let us proceed to calculate the isospin amplitudes $A_{3/2}$ and
$A_{1/2}$ with the help of the factorization hypothesis.
First of all, we assume that final-state interactions were absent
(i.e., $\delta =0$). In this assumption, the transition amplitudes of
$\bar{B}^{0}_{d}\rightarrow D^{+}\pi^{-}$,
$\bar{B}^{0}_{d} \rightarrow D^{0}\pi^{0}$ and
$B^{-}_{u}\rightarrow D^{0}\pi^{-}$ can be expressed in terms
of three topologically different quark-diagram amplitudes: $X$
(color-favored topology), $Y$ (color-suppressed topology) and
$Z$ (annihilation topology) \cite{XL95,Neubert}. Explicitly, we have
\begin{eqnarray}
A_{+-}(\delta =0) & = & ~ X ~ + ~ Z \; ,
\nonumber \\
A_{00}(\delta =0) ~ & = & \frac{Y}{\sqrt{2}} - \frac{Z}{\sqrt{2}} \; ,
\nonumber \\
A_{0-}(\delta =0) & = & ~ X ~ + ~ Y \; ,
%		(6)
\end{eqnarray}
where
\begin{eqnarray}
X & = & \frac{G_{\rm F}}{\sqrt{2}} ~ a^{\rm eff}_1 (V_{cb}V^*_{ud})
\langle \pi^- |(\bar{d}u)^{~}_{\rm V-A}|0\rangle
\langle D^+|(\bar{c}b)^{~}_{\rm V-A}|\bar{B}^0_d\rangle \; ,
\nonumber \\
Y & = & \frac{G_{\rm F}}{\sqrt{2}} ~ a^{\rm eff}_2 (V_{cb}V^*_{ud})
\langle D^0 |(\bar{c}u)^{~}_{\rm V-A}|0\rangle
\langle \pi^-|(\bar{d}b)^{~}_{\rm V-A}|B^-_u\rangle \; ,
\nonumber \\
Z & = & \frac{G_{\rm F}}{\sqrt{2}} ~ a^{\rm eff}_2 (V_{cb}V^*_{ud})
\langle D^+\pi^- |(\bar{c}u)^{~}_{\rm V-A}|0\rangle
\langle 0|(\bar{d}b)^{~}_{\rm V-A}|\bar{B}^0_d\rangle \;
%		(7)
\end{eqnarray}
in the QCD-improved factorization approximation \cite{F}. Here
$a^{\rm eff}_1$ and $a^{\rm eff}_2$ are the effective Wilson coefficients,
$V_{cb}$ and $V_{ud}$ are the relevant Cabibbo-Kobayashi-Maskawa
matrix elements. Now we take final-state interactions into
account (i.e., $\delta \neq 0$). The isospin amplitudes $A_{3/2}$
and $A_{1/2}$ can then be written as
\begin{eqnarray}
A_{3/2} & = & \left ( \frac{X}{3} + \frac{Y}{3} \right )
e^{i\delta_{3/2}} \; ,
\nonumber \\
A_{1/2} & = & \left ( \frac{\sqrt{2}X}{3} - \frac{Y}{3\sqrt{2}}
+ \frac{Z}{\sqrt{2}} \right ) e^{i\delta_{1/2}} \; ,
%		(8)
\end{eqnarray}
where $\delta_{3/2}$ and $\delta_{1/2}$ represent the strong phases
of $I=3/2$ and $I=1/2$ isospin configurations, respectively.
Note that $\delta_{3/2} - \delta_{1/2} = \delta$ holds by definition.
Substituting Eq. (8) into Eq. (3) and taking $\delta_{3/2}=\delta_{1/2}$,
we are able to reproduce Eq. (6).

\vspace{0.3cm}

In comparison with $X$ and $Y$, the annihilation topology $Z$ is
expected to have significant form-factor suppression \cite{Xing96}.
Therefore we neglect $Z$ and obtain
\begin{equation}
r \; =\; \sqrt{2} ~ \frac{X + Y}{2X -Y} \; =\;
\sqrt{2} ~ \frac{a^{\rm eff}_1 + \zeta a^{\rm eff}_2}
{2a^{\rm eff}_1 - \zeta a^{\rm eff}_2}
%		(9)
\end{equation}
from Eqs. (7) and (8), where
\begin{eqnarray}
\zeta & \equiv & \frac{\langle D^0 |(\bar{c}u)^{~}_{\rm V-A}|0\rangle
\langle \pi^-|(\bar{d}b)^{~}_{\rm V-A}|B^-_u\rangle}
{\langle \pi^- |(\bar{d}u)^{~}_{\rm V-A}|0\rangle
\langle D^+|(\bar{c}b)^{~}_{\rm V-A}|\bar{B}^0_d\rangle}
\nonumber \\
& = & \frac{(m^2_B - m^2_\pi) ~ f_D ~ F^{B\rightarrow \pi}_0(m^2_D)}
{(m^2_B - m^2_D) ~ f_\pi ~ F^{B\rightarrow D}_0(m^2_\pi)} \;\; .
%		(10)
\end{eqnarray}
Using $r \approx 1.0$ obtained in Eq. (5) and $\zeta \approx 0.9$ given in
Ref. \cite{Neubert}, we can determine the ratio of the effective Wilson
coefficients $a^{\rm eff}_1$ and $a^{\rm eff}_2$ with the help
of Eq. (9):
\begin{equation}
\frac{a^{\rm eff}_2}{a^{\rm eff}_1} \; =\;
\frac{\sqrt{2}}{\zeta} \cdot \frac{\sqrt{2} ~ r - 1}{r + \sqrt{2}}
\;\; \approx \; 0.27 \; .
%		(11)
\end{equation}
This result is in good agreement with the theoretical expectation \cite{F}.

\vspace{0.3cm}

Next we turn to the decay modes $\bar{B}^0_d \rightarrow D^{*+} \pi^-$,
$\bar{B}^0_d \rightarrow D^{*0} \pi^0$ and
$B^-_u \rightarrow D^{*0}\pi^-$. The Belle and CLEO Collaborations have
recently reported the evidence for the color-suppressed transition
$\bar{B}^0_d \rightarrow D^{*0} \pi^0$, from which a preliminary
value of the branching ratio can be obtained \cite{Belle,CLEO}:
\begin{equation}
\tilde{\cal B}_{00} \; = \; \left \{ \matrix{
(1.5^{+0.6+0.3}_{-0.5-0.4}) \cdot 10^{-4} ~~~~ ({\rm Belle}) \; ,
\;\;\;\;\;\;\;\;\;\; \cr
(2.0 \pm 0.5 \pm 0.7) \cdot 10^{-4} ~~~~ ({\rm CLEO}) \; .} \right .
%		(12)
\end{equation}
In contrast, the color-favored decay modes
$\bar{B}^0_d \rightarrow D^{*+} \pi^-$ and $B^-_u \rightarrow D^{*0}\pi^-$
have the following branching ratios \cite{PDG}:
\begin{eqnarray}
\tilde{\cal B}_{+-} & = & \left (2.76 \pm 0.21 \right ) \cdot 10^{-3} \; ,
\nonumber \\
\tilde{\cal B}_{0-} & = & \left (4.6 \pm 0.4 \right ) \cdot 10^{-3} \; .
%		(13)
\end{eqnarray}
As $B\rightarrow D^*\pi$ decays have the same isospin configurations
as $B\rightarrow D\pi$, one may carry out an analogous analysis
to determine the ratio of $I=3/2$ and $I=1/2$ isospin amplitudes
(both its magnitude $\tilde{r}$ and its phase $\tilde{\delta}$) for
the former. We obtain
\begin{eqnarray}
\tilde{r} & \approx & \left \{ \matrix{
0.98 ~~~~ ({\rm Belle}) \; , \;\; \cr
0.97 ~~~~ ({\rm CLEO}) \; ;} \right .
\nonumber \\
\tilde{\delta} & \approx & \left \{ \matrix{
19^\circ ~~~~ ({\rm Belle}) \; , \;\; \cr
25^\circ ~~~~ ({\rm CLEO}) \; ,} \right .
%		(14)
\end{eqnarray}
using the central values of $\tilde{\cal B}_{+-}$, $\tilde{\cal B}_{00}$,
$\tilde{\cal B}_{0-}$ and $\kappa$. We see that the magnitude of
final-state interactions in $B\rightarrow D^*\pi$ decays might be comparable
with that in $B\rightarrow D\pi$ decays. A more precise measurement of
$\bar{B}^0_d\rightarrow D^{*0}\pi^0$ will narrow the error bar associated
with its branching ratio $\tilde{\cal B}_{00}$ and allow us to extract
the value of $\tilde{\delta}$ reliably. The result $\tilde{r}\approx 1$,
similar to $r \approx 1$ for $B\rightarrow D\pi$, indicating that
the two isospin amplitudes have comparable contributions to the
decay modes $\bar{B}^0_d\rightarrow D^{*+}\pi^-$ and
$\bar{B}^0_d\rightarrow D^{*0}\pi^0$.

\vspace{0.3cm}

Applying the factorization approximation to $B\rightarrow D^*\pi$
decays, one may analogously calculate the ratio of the effective Wilson
coefficients $\tilde{a}^{\rm eff}_1$ and $\tilde{a}^{\rm eff}_2$.
The result is
\begin{equation}
\frac{\tilde{a}^{\rm eff}_2}{\tilde{a}^{\rm eff}_1} \; =\;
\frac{\sqrt{2}}{\tilde{\zeta}} \cdot
\frac{\sqrt{2} ~ \tilde{r} - 1}{\tilde{r} + \sqrt{2}}
\;\; \approx \; 0.25 \; ,
%		(15)
\end{equation}
where
\begin{eqnarray}
\tilde{\zeta} & \equiv &
\frac{\langle D^{*0} |(\bar{c}u)^{~}_{\rm V-A}|0\rangle
\langle \pi^-|(\bar{d}b)^{~}_{\rm V-A}|B^-_u\rangle}
{\langle \pi^- |(\bar{d}u)^{~}_{\rm V-A}|0\rangle
\langle D^{*+}|(\bar{c}b)^{~}_{\rm V-A}|\bar{B}^0_d\rangle}
\nonumber \\
& = & \frac{f_{D^*} ~ F^{B\rightarrow \pi}_+(m^2_{D^*})}
{f_\pi ~ A^{B\rightarrow D^*}_0(m^2_\pi)} \; \approx \; 0.9
%		(16)
\end{eqnarray}
has been used \cite{Neubert}. One can see that
the values of $\tilde{a}^{\rm eff}_2/\tilde{a}^{\rm eff}_1$
and $a^{\rm eff}_2/a^{\rm eff}_1$ are consistent with each other.

\vspace{0.3cm}

In summary, we have analyzed the isospin relations for the amplitudes
of $\bar{B}^0_d \rightarrow D^{(*)+}\pi^-$,
$\bar{B}^0_d \rightarrow D^{(*)0}\pi^0$ and
$B^-_u \rightarrow D^{(*)0}\pi^-$ transitions. The strong
phase differences are found to be about $29^\circ$
(or $26^\circ$) and $19^\circ$ (or $25^\circ$), respectively,
in $B\rightarrow D\pi$ and $B\rightarrow D^*\pi$.
We have also applied the factorization hypothesis to the decay modes
under discussion. We find that the value of the factorization parameter
$a^{\rm eff}_2/a^{\rm eff}_1$ extracted from $B\rightarrow D\pi$ is
compatible with that extracted from $B\rightarrow D^*\pi$, and both of
them are in good agreement with the theoretical expectation.
We await more precise measurements of the color-suppressed decay modes
$\bar{B}^0_d \rightarrow D^0\pi^0$ and
$\bar{B}^0_d \rightarrow D^{*0}\pi^0$ at $B$-meson factories, in order
to make a more stringent test of the factorization approximation and
a more accurate analysis of final-state interactions.

\vspace{0.5cm}

I am deeply indebted to H.Y. Cheng for his enlightening
comments on this paper, for correcting a crucial numerical error
in its original version, and for calling my attention to the new
CLEO data. I am also grateful to T.E. Browder for pointing out a
typing error, and to J.F. Sun and D.S. Yang for useful discussions.

\end{document}